\documentclass[journal,draft,onecolumn]{IEEEtran}
\usepackage{times}
\usepackage[final]{graphicx}
\usepackage[reqno]{amsmath}
\usepackage{amsfonts}

\usepackage{times,amsmath,epsfig}
\usepackage{latexsym,amssymb}
\usepackage[center]{caption}
\usepackage{graphicx}         
\usepackage{color}            

\newcommand{\Prob}{\textrm{Pr}}
\newcommand{\beq}{\begin{equation}}
\newcommand{\enq}{\end{equation}}
\newcommand{\beqa}{\begin{eqnarray}}
\newcommand{\enqa}{\end{eqnarray}}
\newcommand{\beqn}{\begin{eqnarray*}}
\newcommand{\enqn}{\end{eqnarray*}}

\newcommand{\no}{\nonumber}

\newtheorem{theorem}{Theorem}

\newtheorem{lemma}[theorem]{Lemma}

\newcommand{\qed}{\hfill $\Box$}

\ifCLASSINFOpdf
\else
\fi
\begin{document}
%
\title{Keys through ARQ}
%
%
%

\author{Mohamed~Abdel Latif,~\IEEEmembership{Student Member,~IEEE,}
        Ahmed~Sultan,~\IEEEmembership{Member,~IEEE,}
        and~Hesham~El Gamal,~\IEEEmembership{Senior Member,~IEEE}

\thanks{Mohamed~Abdel Latif and Ahmed~Sultan are with the Wireless Intelligent Networks Center (WINC), Nile University, Cairo, Egypt(mohamed.abdelghany@nileu.edu.eg, asultan@nileuniversity.edu.eg). Hesham~El~Gamal is with the ECE Department, The Ohio State University, Columbus, OH, USA (helgamal@ece.osu.edu). This work was partly funded by the National Telecom Regulatory Authority (NTRA) of Egypt and the National Science Foundation (NSF).}}

%
%

\markboth{Submitted to the IEEE Transactions on Information Theory}%
{Shell \MakeLowercase{\textit{et al.}}: Bare Demo of IEEEtran.cls for Journals}
%



\maketitle

\begin{abstract}
This paper develops a novel framework for sharing secret keys using the well-known \textbf{A}utomatic \textbf{R}epeat re\textbf{Q}uest (\textbf{ARQ}) protocol. The proposed key sharing protocol does not assume any prior knowledge about the channel state information (CSI), but, harnesses the available opportunistic secrecy gains using only the one bit feedback, in the form of ACK/NACK. The distribution of key bits among multiple ARQ epochs, in our approach, allows for mitigating the secrecy outage phenomenon observed in earlier works. We characterize the information theoretic limits of the proposed scheme, under different assumptions on the channel spatial and temporal correlation function, and develop low complexity explicit implementations. Our analysis reveals a novel role of ``dumb antennas" in overcoming the negative impact of spatial correlation, between the legitimate and eavesdropper channels, on the achievable secrecy rates. We further develop an adaptive rate allocation policy which achieves higher secrecy rates by exploiting the channel temporal correlation. Finally, our theoretical claims are validated by numerical results that establish the achievability of non-zero secrecy rates even when the eavesdropper channel is less noisy, on the average, than the legitimate channel.
\end{abstract}



%
\IEEEpeerreviewmaketitle

\section{Introduction}
%
%
%
%
The recent flurry of interest on wireless physical layer
secrecy is inspired by Wyner's pioneering work on the wiretap channel. Under the assumption that the eavesdropper channel is a degraded version of the legitimate channel, Wyner showed in~~\cite{Wyner1,Wyner2} that perfectly secure communication is possible by hiding the message
in the additional noise level seen by the eavesdropper. The effect of fading on the secrecy capacity was studied later. In particular, by appropriately distributing the message across different fading realizations, it was shown that the multi-user diversity gain can be harnessed to enhance the secrecy capacity, e.g.~\cite{fading,NOCSI}. More recently, the authors of~\cite{Poor} proposed using the well-known Hybrid ARQ protocol to facilitate the exchange of secure messages over fading channels. This paper extends this line of work by developing a novel ARQ-based approach for secret key sharing between two legitimate users (Alice and Bob), communicating over a wireless channel, in the presence of a passive eavesdropper (Eve). The shared key can then be used to secure any future message transmission.

One innovative aspect of our framework is the distribution of key bits over an asymptotically large number of ARQ epochs. This approach allows for overcoming the secrecy outage phenomenon observed in~\cite{Poor} at the expense of increased delay. In this setup, we characterize the fundamental information theoretic limits on the maximum achievable key rate; subject to a perfect secrecy constraint. Our information theoretic analysis inspires the design of explicit ARQ protocols that attain an excellent throughput-delay-secrecy tradeoff with a realizable coding/decoding complexity. It also reveals the negative impact of spatial correlation on the achievable key rate. This problem is mitigated via the efficient use of dumb antennas which is shown to effectively {\em decorrelate} the legitimate and eavesdropper channels in the asymptotic limit of a large number of transmit antennas. Moreover, we propose a greedy rate adaptation algorithm capable of transforming the temporal correlation in the legitimate channel into additional gains in the secrecy rate. In a nutshell, our results demonstrate the achievability of non-zero perfectly secure key rate over fading channels by opportunistically exploiting the ARQ feedback (even when the eavesdropper channel is {\em less noisy}, {\em on the average}, than the main channel).

The rest of this paper is organized as follows. Our system model is detailed in Section~\ref{sec:sysmodel}. Section~\ref{sec:indep} develops the main results for the spatially independent block fading model. In~Section~\ref{sec:corr}, we extend our analysis to spatially and temporally correlated channels, whereas numerical results that validate our theoretical claims are presented in Section~\ref{sec:numeric}. Finally, Section~\ref{sec:conc} offers some concluding remarks and our proofs are collected in the Appendices to enhance the flow of the paper.

\section{System Model}\label{sec:sysmodel}

\begin{figure}
\centering
\input{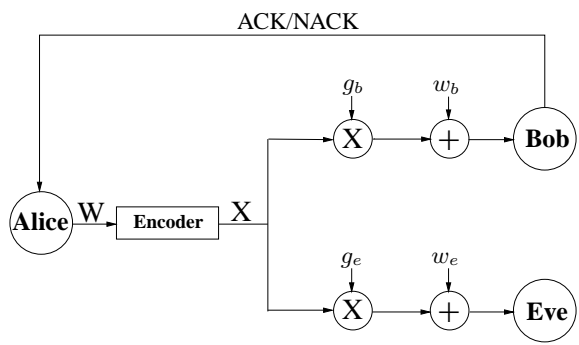}
\caption{System model involves a legitimate receiver, Bob, with a
feedback channel to the sender, Alice. Eve is a passive
eavesdropper. We assume block fading channels that are independent
of each other. \label{model1}}
\end{figure}

Our model, shown in Figure~\ref{model1}, assumes one transmitter (Alice), one legitimate receiver
(Bob) and one passive eavesdropper (Eve). We adopt a block fading model in which the channel is
assumed to be fixed over one coherence interval and changes from one interval to the next. In order to obtain rigorous
information theoretic results, we consider the scenario of
asymptotically large coherence intervals and allow for sharing the
secret key across an asymptotically large number of those intervals. The
finite delay case will be considered as well. In
any particular interval, the signals received by Bob and Eve are
respectively given by,
\begin{eqnarray}
y(i,j)&=&g_b(i)\, x(i,j)+w_b(i,j),\\
z(i,j)&=&g_e(i)\, x(i,j)+w_e(i,j),
\end{eqnarray}
\noindent where $x(i,j)$ is the $\it {j}^{th}$ transmitted symbol
in the $\it {i}^{th}$ block, $y(i,j)$ is the $\it {j}^{th}$
received symbol by Bob in the $\it {i}^{th}$ block, $z(i,j)$ is
the $\it {j}^{th}$ received symbol by Eve in the $\it {i}^{th}$
block, $g_{b}(i)$ and $g_{e}(i)$ are the complex block channel
gains from Alice to Bob and Eve, respectively. The channel gains can also be written as
\begin{eqnarray}
g_{b}(i) = \sqrt{h_{b}(i)}\exp(j\theta_b(i))\\
g_{e}(i) = \sqrt{h_e(i)}\exp(j\theta_e(i)),
\end{eqnarray}
where $\theta_b(i)$ and $\theta_e(i)$, the phase shifts at Bob and Eve respectively, are assumed to be independent in \textbf{all} considered scenarios. Moreover, $w_{b}(i,j)$ and $w_{e}(i,j)$ are zero-mean, unit variance white complex Gaussian noise coefficients at Bob and Eve, respectively. We do not assume any prior knowledge about the
channel state information at Alice. Bob, however, is assumed to
know $g_b(i)$ and Eve is assumed to know both $g_b(i)$ and
$g_e(i)$ {\em a-priori}. We impose the following short-term average power
constraint
\begin{eqnarray}
 {\mathbb E} \left(|x(i,j)|^2\right)\leq \bar{P}.
\end{eqnarray}

Our model only allows for one bit of ARQ feedback from Bob to Alice. Each ARQ epoch is assumed to be contained in one
coherence interval (i.e., fixed channel gains) and that different
epochs correspond to different coherence intervals. The transmitted
packets are assumed to carry a perfect error detection mechanism
that Bob (and Eve) can use to determine whether the packet has been
received correctly or not. Based on the error check, Bob sends
back to Alice an ACK/NACK bit, through a public and error-free
feedback channel. Eve is assumed to be passive (i.e., can not
transmit); an assumption which can be justified in several
practical settings. To minimize Bob's receiver complexity, we
adopt the memoryless decoding assumption implying that frames
received in error are discarded and not used to aid in future
decoding attempts.

\section{Secrecy via ARQ}\label{sec:indep}
Our main results are first derived for the scenario where $h_b$ and $h_e$ vary \textbf{independently} from one block to another according to a joint distribution $f\left(h_b,h_e\right)$. The impact of temporal correlation on the performance of our secret key sharing protocols will be investigated in the next section.

\subsection{Information Theoretic Foundation}
In our setup, Alice wishes to share a secret key $W \in {\mathcal
W}=\{1,2,\cdots ,M\}$ with Bob. To transmit this key, Alice and Bob use an
$(M,m)$ code consisting of : 1) a stochastic encoder $f_m(.)$ at Alice
that maps the key $w$ to a codeword $x^m \in {\mathcal X}^{m}$, 2) a
decoding function $\phi$: ${\mathcal Y}^{m}\rightarrow {\mathcal W}$
which is used by Bob to recover the key. The codeword is partitioned
into $a$ blocks, each one corresponds to one ARQ-epoch and contains $n_1$ symbols where $m = a\,n_1$. For now, we focus on the asymptotic scenario where
$a\rightarrow\infty$ and $n_1\rightarrow\infty$.

Alice starts with a random selection of the first block of $n_1$
symbols. Upon reception, Bob attempts to decode this block. If
successful, it sends an ACK bit to Alice who moves ahead and
makes a random choice of the second $n_1$ and sends it to Bob.
Here, Alice must make sure that the concatenation of the two
blocks belong to a valid codeword. As shown in the sequel, this
constraint is easily satisfied. If an error was detected, then Bob
sends a NACK bit to Alice. To simplify the analysis, we assume that the error
detection mechanism is perfect which is justified in the asymptotic scenario
$n_1\rightarrow\infty$. In this case, Alice {\bf replaces} the
first block of $n_1$ symbols with another randomly chosen block
and transmits it. The process then repeats until Alice and Bob
agree on a sequence of $a$ blocks, each of length $n_1$ symbols,
corresponding to the key.

The code construction must allow for reliable decoding at Bob while
hiding the key from Eve. It is clear that the proposed protocol
exploits the error detection mechanism to make sure that both Alice and
Bob agree on the key (i.e., ensures reliable decoding). What remains
is the secrecy requirement which is measured by the equivocation
rate $R_e$ defined as the entropy rate of the transmitted key
conditioned on the intercepted ACKs or NACKs and the channel outputs
at Eve, i.e.,
\begin{equation}
R_e ~\overset{\Delta}{=}~ \frac{1}{n} H(W|Z^n,K^b,G_b^b,G_e^b) ~,
\end{equation} where $n$ is the number of symbols transmitted to exchange the key (including the symbols in the discarded blocks due to decoding
errors), $b=a\frac{n}{m}$, $K^b = \{ K(1), \cdots, K(b)\}$ denotes
sequence of ACK/NACK bits, $G_b^b$ and $G_e^b$ are the sequences of
channel coefficients seen by Bob and Eve in the $b$ blocks, and $Z^n
= \{ Z(1), \cdots, Z(n)\}$ denotes Eve's channel outputs in the $n$
symbol intervals. We limit our attention to the perfect secrecy
scenario, which requires the equivocation rate $R_e$ to be
arbitrarily close to the key rate. The secrecy rate $R_s$ is said to
be achievable if for any $\epsilon>0$, there exists a sequence of
codes $(2^{nR_s},m)$ such that for any $m\geq m(\epsilon)$, we have
\begin{equation}\label{secrecy}
R_e ~=~ \frac{1}{n} H(W|Z^n,K^b,G_b^b,G_e^b) ~\geq~ R_{s}-\epsilon
\end{equation}
and the {\bf key rate} for a given input distribution is defined as the maximum achievable perfect secrecy rate with this distribution. The following result characterizes this rate, assuming a Gaussian input distribution

\begin{theorem}\label{thm1}
The key rate for the memoryless ARQ protocol with {\bf
Gaussian inputs} is given by:
\begin{equation}
C_s^{(g)} = \max\limits_{R_0,P\leq \bar{P}}\mathbb{E}\left\{\left[R_0 - \log_2\left(1+h_eP\right)\right]^{+}\mathbb{I}\left(R_0 \leq \log_2\left(1+h_bP\right) \right)\right\}
\label{sec_cap_general}
\end{equation}
where $[x]^+=\max(0,x)$ and $\mathbb{I}(x) = 1$ if $x$ is true and $0$ otherwise.
For the special case of spatially independent fading, i.e. $f\left(h_b,h_e\right) = f(h_b)f(h_e)$) the above expression simplifies to
\begin{equation}
C_s^{(i)} =  \max\limits_{R_0,P\leq \bar{P}}\{\textrm{Pr}(R_0 \leq \log_2(1+h_bP)){\mathbb E}[R_0 - \log_2(1 + h_eP)]^{+}\}
\label{sec_cap_special}
\end{equation}
\end{theorem}

A few remarks are now in order
\begin{enumerate}
\item It is clear from (\ref{sec_cap_general}) that a positive
    secret key rate is achievable under very mild conditions on the
    channels experienced by Bob and Eve. More precisely, unlike the
    approach proposed in~\cite{Poor}, Theorem~\ref{thm1} establishes
    the achievability of a positive perfect secrecy rate by
    appropriately exploiting the ARQ feedback even when Eve's average
    SNR is higher than that of Bob.

\item Theorem~\ref{thm1} characterizes the fundamental limit on secret key sharing and not
    message transmission. The difference between the two scenarios
    stems from the fact that the message is known to Alice {\bf
    before} starting the transmission of the first block, whereas Alice
    and Bob can defer the agreement on the key till the last
    successfully decoded block. This observation was exploited by our
    approach in making Eve's observations of the frames discarded by
    Bob, due to failure in decoding, useless.
\item It is intuitively pleasing that the secrecy key rate in
    (\ref{sec_cap_special}) is the product of the probability of success at
    Bob and the expected value of the additional mutual information
    gleaned by Bob, as compared to Eve, in those successfully decoded
    frames.
\item We stress the fact that our approach does not require any prior knowledge about the
    channel state information. The only assumption is that the public feedback channel is error-free, authenticated, and only accessible by Bob.
\item The achievability of (\ref{sec_cap_general}) hinges on a random
    binning argument which only establishes the existence of a coding
    scheme that achieves the desired rate. Our result, however,
    stops short of explicitly finding such optimal coding scheme and
    characterizing its encoding/decoding complexity. This observation
    motivates the development of the explicit secrecy coding schemes in Section~\ref{sec:imp}.
\end{enumerate}

\subsection{Explicit Secrecy Coding Schemes}\label{sec:imp}
This section develops explicit secrecy coding schemes that allow
for sharing keys using the underlying memoryless ARQ protocol with realizable encoding/decoding complexity and delay. We
proceed in three steps. The first step replaces the random binning
construction, used in the achievability proof of
Theorem~\ref{thm1}, with an explicit coset coding scheme for the
erasure-wiretap channel. This erasure-wiretap
channel is created by the ACK/NACK feedback and accounts for the
computational complexity available to Eve. In the second step, we
limit the decoding delay by distributing the key bits over only a
finite number of ARQ frames. Finally, we replace the capacity
achieving Gaussian channel code with practical coding schemes in
the third step. Overall, our three-step approach allows for a nice
performance-vs-complexity tradeoff.

The perfect secrecy requirement used in the information theoretic
analysis does not impose any limits on Eve's decoding complexity. The
idea now is to exploit the finite complexity available at Eve in
simplifying the secrecy coding scheme. To illustrate the idea, let's
first assume that Eve can only afford maximum likelihood (ML)
decoding. Hence, successful decoding at Eve is only possible when

\begin{equation}
R_0\leq \log_2(1+h_e P),
\end{equation}

for a given transmit power level $P$. Now, using the idealized
error detection mechanism, Eve will be able to identify and {\bf erase}
the frames decoded in error resulting in an {\bf erasure wiretap channel model}. In practice, Eve may be able to go beyond the performance of the ML
decoder. For example, Eve can generate a list of candidate codewords
and then use the error detection mechanism, or other means, to
identify the correct one. In our setup, we quantify the
computational complexity of Eve by the amount of side information
$R_{\rm c}$ bits per channel use offered to it by a Genie. With this side
information, the erasure probability at Eve is given by

\begin{equation}
\epsilon=\textrm{Pr}\left(R_0-R_{\rm c}> \log_2(1+h_eP)\right),\label{rc}
\end{equation}
since now the channel has to supply only enough mutual information
to close the gap between the transmission rate $R_0$ and the side
information $R_{\rm c}$. The ML performance can be obtained as a
special case of (\ref{rc}) by setting $R_{\rm c}=0$.

\begin{figure}
\centering
\input{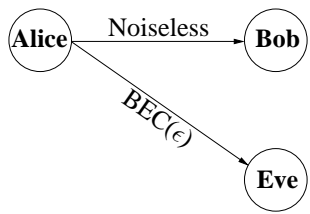}
\caption{Erasure-wiretap channel equivalent model. \label{BEC}}
\end{figure}

It is now clear that using this idea we have transformed our ARQ
channel into an erasure-wiretap channel, as in Figure~\ref{BEC}. In this equivalent model,
we have a noiseless link between Alice and Bob, ensured by the
idealized error detection algorithm, and an erasure channel between
Alice and Eve. The following result characterizes the achievable
performance over this channel

\begin{lemma}\label{erasure1}
The secrecy capacity for the equivalent erasure-wiretap channel is
\begin{eqnarray}
C_e &=&  \max\limits_{R_0, P\leq \bar{P}} \left\{R_0 \mathbb{E}\left[\mathbb{I}\left(\left(R_0 \leq \log_2(1+h_bP)\right).\left(R_0 - R_c \geq \log_2(1+h_eP)\right)\right)\right]\right\}\nonumber \\
    &=&  \max\limits_{R_0, P\leq \bar{P}}  \{R_0\textrm{Pr}(R_0 \leq
\log_2(1+h_bP),R_0-R_{\rm c} > \log_2(1+h_eP))\}
\end{eqnarray}
In the case of spatially independent channels, the above expression reduces to
\begin{equation}
C_e =  \max\limits_{R_0, P\leq \bar{P}} \{R_0~\textrm{Pr}(R_0 \leq \log_2(1+h_bP))~\textrm{Pr}(R_0-R_{\rm c} > \log_2(1+h_eP))\}\label{sp_erasure}
\end{equation}
\end{lemma}

The proof follows from the classical result on the erasure-wiretap
channel~\cite{Wyner2}. It is intuitively
appealing that the expression in (\ref{sp_erasure}) is simply
the product of the transmission rate per channel use, the
probability of successful decoding at Bob, and the probability of
erasure at Eve. The main advantage of this equivalent model is
that it lends itself to the explicit coset LDPC coding scheme
constructed in~\cite{LDPC1,LDPC2,LDPC3}. In summary, our first low
complexity construction is a concatenated coding scheme where the
outer code is a coset LDPC for secrecy and the inner one is a
capacity achieving Gaussian code. {\bf The underlying memoryless ARQ is
used to create the erasure-wiretap channel matched to this
concatenated coding scheme}.

The second step is to limit the decoding delay resulting from the
distribution of key bits over an asymptotically large number of
ARQ blocks in the previous approach. To avoid this problem, we
limit the number of ARQ frames used by the key to a finite number
$k$. The implication for this choice is a non-vanishing value for
secrecy outage probability. For example, if we encode the message as the syndrome of the rate $(k-1)/k$ parity check code then Eve will be completely blind about the key if {\em
at least} one of the $k$ ARQ frames is
erased~\cite{LDPC1,LDPC2,LDPC3} (Here the distilled key is the modulo-$2$ sum of the key parts received correctly).
The secrecy outage probability, assuming spatially independent channels, is therefore

\begin{equation}
P_{\rm out}= {\rm Pr}\left(\min\limits_{j\in\{1,...,k\}} \log_2
(1+h_{e}(j)P)> R_0 - R_{\rm c}\right),
\end{equation}
where $h_e(1)$,...,$h_e(k)$ are i.i.d. random variables drawn
according to the marginal distribution of Eve's channel. Assuming a Rayleigh
fading distribution, we get

\begin{equation}
P_{\rm out}=\exp\left(-\frac{k}{P}\left[2^{R_0-R_{\rm
c}}-1\right]\right) \label{pout}.
\end{equation}

Under the same assumption, it is straightforward to
see that the average number of Bernoulli trials required to transfer
$k$ ARQ frames successfully to Bob is given by

\begin{equation}
N_0=k\exp\left(\frac{2^{R_0}-1}{P}\right), \label{avg_number}
\end{equation}

\noindent resulting in a key rate

\begin{equation}
R_k=\frac{R_0}{N_0}=\frac{R_0}{k}\exp\left(-\frac{2^{R_0}-1}{P}\right).
\end{equation}

Therefore, for a given $R_{\rm c}$ and $P$, one can obtain a
tradeoff between $P_{\rm out}$ and $R_k$ by varying $R_0$. Our
third, and final, step is to relax the assumption of a capacity
achieving inner code. Section~\ref{sec:numeric} reports numerical results with practical coding schemes,
including uncoded transmission, with a finite
frame length $n_1$. Overall, these results demonstrate the ability of the proposed protocols to achieve near-optimal key rates, under very mild assumptions, with realizable encoding/decoding complexity and bounded delay.

\section{Correlated Fading}\label{sec:corr}
\subsection{Dumb Antennas for Secrecy}
One of the important insights revealed by Theorem~\ref{thm1} is the negative relation between the achievable key rate and the spatial correlation between the main and eavesdropper channels. In fact, one can easily verify that the key rate collapses to zero in the fully correlated case (i.e., $h_b=h_e$ with probability one) independent of the marginal distribution of $h_b$. In this section, we propose a solution to this problem based on a novel utilization of ``dumb antennas." The concept of dumb antennas was introduced in~\cite{Dumb} as a means to create artificial channel fluctuations in slow fading environments. These fluctuations are used to harness opportunistic performance gains in multi-user cellular networks. As indicated by the name, one of the attractive features of this approach is that the receiver(s) can be oblivious to the presence of multiple transmit antennas~\cite{Dumb}. We use dumb transmit antennas to de-correlate the main and eavesdropper channels as follows. Alice is equipped with $N$ transmit antennas, whereas both Bob and Eve will still have only one receive antenna. In order to simplify the presentation, we focus on the case of the symmetric fully correlated line of sight channels; whereby the magnitudes of the channel gains are all equal to one. The rest of our modeling assumption remains as detailed in~Section~\ref{sec:sysmodel}. The same data stream is transmitted from the $N$ transmitted after applying an i.i.d uniform phase to each of the $N$ signals. Also, Bob is assumed to perturb its location in each ARQ frame resulting in a random and independent phase shift (from that experienced by Eve). Our multiple transmit antenna scenario, therefore, reduces to a single antenna fading wiretap channel with the following {\bf equivalent} channel gains

\begin{eqnarray}
g_b^{eq} = \sum\limits_{n=1}^N \left( \frac{1}{\sqrt{N}} \exp(\theta_{iR} + \theta_{iB})\right)\\
g_e^{eq} = \sum\limits_{n=1}^N \left( \frac{1}{\sqrt{N}} \exp(\theta_{iR}+\theta_{iE})\right),
\end{eqnarray}

\noindent where $\theta_{iB}$, $\theta_{iE}$, and $\theta_{iR}$ are i.i.d. and uniform over $[-\pi,\pi]$ that remain fixed from one ARQ frame and change randomly from one frame to the next. One can now easily see that as $N$ increases, the marginal distribution of each equivalent channel gain approaches a zero-mean complex Gaussian with unit variance (by the Central Limit Theorem (CLT)~\cite{Prob}). It is worth noting that the correlation coefficient between the two channels' equivalent power gains depends on the instantaneous channels' phases $\theta_{iB}$'s and $\theta_{iE}$'s for $i=1,\ldots,N$. It can be easily shown that, in the limit of $N\rightarrow\infty$, this correlation coefficient between the two channels power gains converges in a mean-square sense to zero (please refer to Appendix B for the proof). Therefore, in the asymptotic limit of a large $N$, our dumb antennas approach has successfully transformed our fully correlated line of sight channel into a symmetric and {\bf spatially independent} Rayleigh wiretap channel; whose secrecy capacity (assuming Gaussian inputs) is reported in Theorem~\ref{thm1}. The numerical results reported in the sequel demonstrate that this result is not limited to line of sight channels, and that this asymptotic behavior can be observed for a relatively small number of transmit antennas.

\subsection{Temporal Correlation}
Thus far, we have assumed that the channel gains affecting different frames are independent. This assumption renders optimal the stationary rate allocation strategy of Theorem~\ref{thm1}. In this section, we relax this assumption by introducing temporal correlation between the channel gains experienced by successive frames. Assuming high temporal correlation and if a stationary rate strategy is employed and it is less than Eve's channel capacity, all the information transmitted will be leaked to Eve. On the other hand, if the rate is much less than Bob's channel capacity, additional gains in the secrecy capacity will not be harnessed. Hence, we are going to employ a \textbf{rate adaptation} strategy in which the optimal rate used in each frame is determined based on the past history of ACK/NACK feedbacks and the rates used in previous blocks. More specifically, following in the footsteps of~\cite{RA_Phil}, the optimal rate allocation policy can be formulated as follows (assuming a short term average power constraint $P$ and a Gaussian input distribution).

\begin{eqnarray}\label{rate-policy}
R_t &=& \arg \max \limits_{R_t}{\left\{\left(C_{s,t} + \sum\limits_{k=t+1}^{\infty}C_{s,k}\right) \Big{|} \mathbf{R}_{t-1}, \mathbf{K}_{t-1}\right\}},
\end{eqnarray}
where
\begin{eqnarray}
C_{s,t} &=& \textrm{Pr}(R_t \leq \log_2(1+h_{b, t}P)){\mathbb E}_{h_e} [ R_t - \log_2 (1+h_{e} P) ]^{+}, \no
\end{eqnarray}

\noindent where $\mathbf{R}_{t-1} = \left[R_0,\cdots,R_{t-1}\right]$ is the vector of previous transmission rates and $\mathbf{K}_{t-1} = \left[K_0,\cdots,K_{t-1}\right]$ is the vector of previously received ACKs and NACKs. The basic idea is that, after frame $(t-1)$, the posteriori distribution of $h_b$ is updated using $\mathbf{R}_{t-1}$ and $\mathbf{K}_{t-1}$. The expected secrecy rate, in future transmissions, is then maximized based on this updated distribution. It is worth noting that the above expression assumes {\bf no spatial correlation} between $h_e$ and $h_b$. This assumption represents the worst case scenario since it prevents Alice from learning the channel gains impairing Eve through the ARQ feedback. Since the channel gain is not observed directly, but through an indicator in the form of ARQ feedback, the optimal rate assignment, when the channel is Markovian, is a Partially Observable Markov Decision Process (POMDP). The solution of this POMDP is computationally intractable except for trivial cases. This motivates the following greedy rate allocation policy
\begin{equation}
R_t = \arg \max \limits_{R_t}{\left\{C_{s,t} \Big{|} \mathbf{R}_{t-1}, \mathbf{K}_{t-1}\right\}}
\end{equation}

Interestingly, the numerical results reported in the following section demonstrate the ability of this simple strategy to harness significant performance gains in first order Markov channels. Note that the performance of {\bf any} rate allocation policy can be upperbounded by the ergodic capacity with transmitter CSI (and short term average power constraint $P$), i.e.,

\begin{eqnarray}\label{rate-ergodic}
C_{er} &=& {\mathbb E}_{h_e,h_b} [\log_2 (1+h_{b} P)- \log_2 (1+h_{e} P) ]^{+},
\end{eqnarray}

\noindent which is achieved by the optimal rate allocation policy $R_t=\log_2 (1+h_{b,t} P)$. In fact, one can view the rate assignment policy of (\ref{rate-policy}) as an attempt to approach the rate of (\ref{rate-ergodic}) by using the ARQ feedback to obtain a better estimate of $h_{b,t}$ after each fading block.

\section{Numerical Results}\label{sec:numeric}
Throughout this section, we focus on the symmetric scenario, where the average SNRs experienced by both Bob and Eve are the same, i.e., ${\mathbb E}\left(h_b\right)={\mathbb E}\left(h_e\right)$ = 1. We further assume Rayleigh fading channels, for
both Bob and Eve. Assuming spatially and temporally independent channels, the achievable secrecy rate in
(\ref{sec_cap_special}) becomes
\begin{equation}
C_s= \max\limits_{R_0} \exp\left(-\frac{2^{R_0}-1}{P}\right). \left\{R_0-\frac{\exp\left(1/P\right)}{\log_e\left(2\right)}\left[E_{\rm i}\left(1/P\right)-E_{\rm i}\left(2^{R_0}/P\right)\right]\right\}
\end{equation}
\noindent where $E_{\rm i}\left(x\right)=\int_{x}^{\infty}\exp\left(-t\right)/t\,dt$.

\begin{figure}
\centering
\includegraphics[width=1\textwidth]{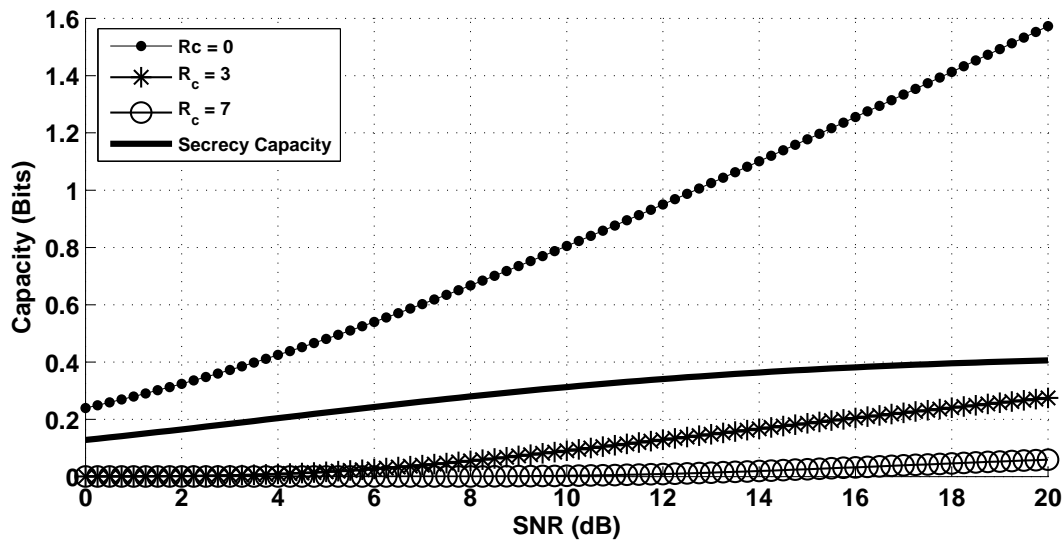}
\caption{$C_s$ and $C_e$ against SNR for $R_c =\left(0,3,7\right)$. \label{fig1}}
\end{figure}

Figure~\ref{fig1} gives the variation of $C_s$ and $C_e$ with SNR
under different constraints on the decoding capabilities of Eve, captured by the genie-given side information, $R_c$. It is clear from the figure that $C_e$ can be greater than $C_s$ for certain $R_c$ and SNR values. For
instance, in the case of $R_c=0$, a
packet received in error at Eve will be discarded {\bf without any further attempts at
decoding}. Therefore, the instantaneous secrecy rate becomes $R_0$, which is
larger than that used in (\ref{sec_cap_special}) $C_s(i)= R_0 -
\log_2(1+h_e(i)P)$ where $C_s(i), h_e(i)$ are the instantaneous
secrecy rate, and Eve's channel power gain, respectively.
Averaging over all fading realizations, we get a greater $C_e$ than
$C_s$. It is worth noting that, under the assumptions of the
symmetric scenario and the Rayleigh fading model, the scheme
proposed in~\cite{Poor} is not able to achieve any positive
secrecy rate (i.e., probability of secrecy outage is one).

\begin{figure}
\centering
\includegraphics[width=1\textwidth]{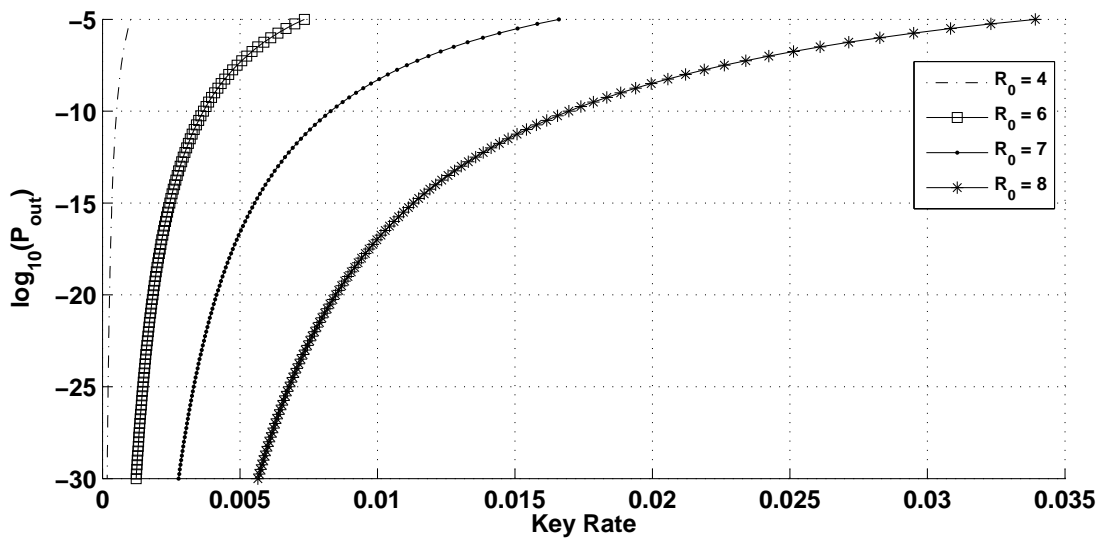}
\caption{Outage probability against key rate for $R_c=2$, $R_o=4$, $6$, $7$ and $8$, and an average SNR of $30$ dB. \label{fig2}}
\end{figure}

\begin{figure}
\centering
\includegraphics[width=1\textwidth]{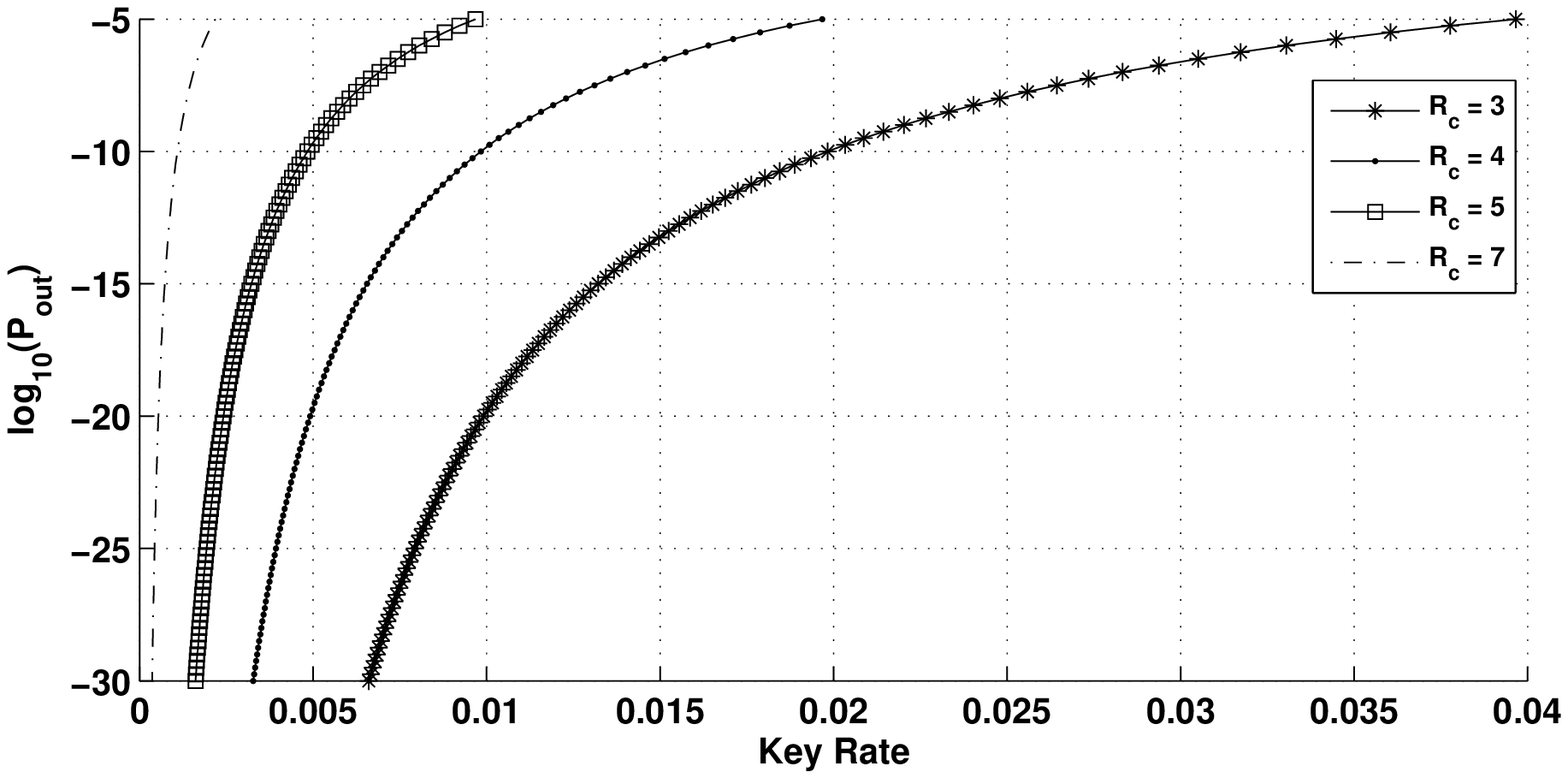}
\caption{Outage probability against key rate for $R_0=10$, $R_c=3$, $4$, $5$ and $7$, and  an average SNR of $30$ dB. \label{fig3}}
\end{figure}

\begin{figure}
\centering
\includegraphics[width=1\textwidth]{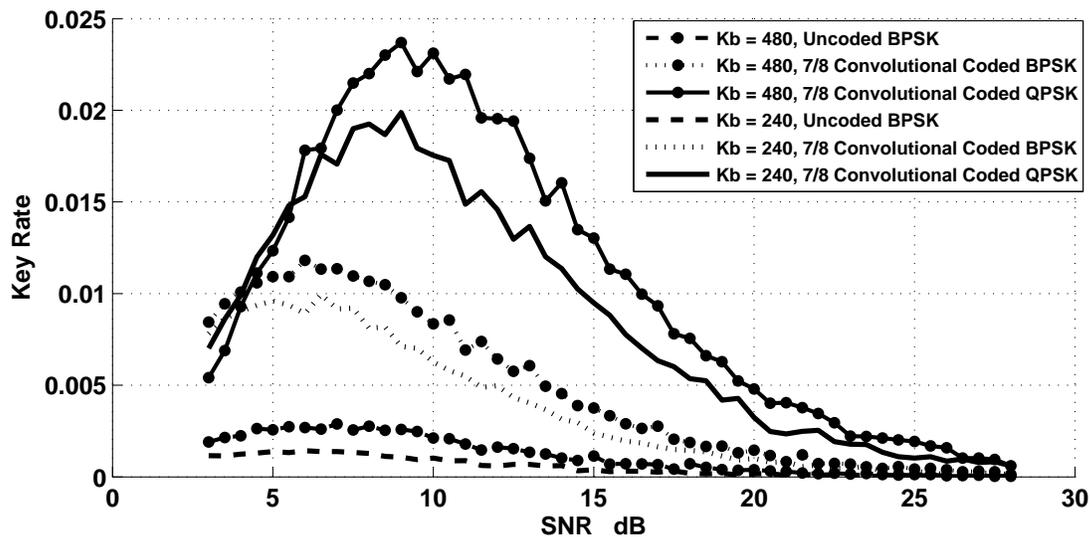}
\caption{The key rates required to obtain an outage of $10^{-10}$
against SNR for different packet sizes, $K_b=240$ and $480$ bits,
and different modulation schemes: uncoded BPSK, coded BPSK, and
coded QPSK. \label{fig4}}
\end{figure}

Next, we turn our attention to the delay-limited coding
constructions proposed in Section~\ref{sec:imp}.
Figures~\ref{fig2}~and~\ref{fig3} show, for different $R_0$ and
$R_{\rm c}$, the tradeoff between the secrecy outage probability
and key rate for the proposed rate $(k-1)/k$ coset secrecy
coding scheme assuming an optimal inner Gaussian channel coding.
Figure~\ref{fig2} gives the key rate corresponding to a desired
secrecy outage probability, given some values for $R_0$ and $R_c$. Figure~\ref{fig3}, on the other hand, quantifies the reduction in key rate, corresponding to a certain outage probability, as $R_c$
increases. In Figure~\ref{fig4}, we relax the optimal channel
coding assumption and plot key rates for practical coding schemes
and finite frame lengthes (i.e., finite $n_1$). The
code used in the simulation is a punctured convolutional code
derived from a basic $1/2$ code with a constraint length of $7$
and generator polynomials $133$ and $171$ (in octal). We assume
that Eve is genie-aided and can correct an additional $50$
erroneous symbols (beyond the error correction capability of the
channel code). From the figure, we see that the key rate increases
with increasing SNR and then drops after reaching a peak value.
Note that the transmission rate is fixed and independent of the
SNR. Therefore, a low SNR means more transmissions to Bob and a consequent
low key rate. As the SNR increases, while keeping the transmission
rate fixed, the key rate increases. However, increasing the SNR also means
an increased ability of Eve to correctly decode the
codeword-carrying packets. This explains why the key rate curves
peak and then decay with SNR. In practice, one can always operate at the optimal value of the SNR by adjusting the transmit power level. We also observe that for a certain modulation and channel coding scheme, decreasing the packet size
in bits lowers the key rate. Reducing the packet size increases
the probability of correct decoding by Bob and, thus, decreases
the number of transmissions. However, it also increases the
probability of correct decoding by Eve and the overall effect is a
decreased key rate.

\begin{figure}
\centering
\includegraphics[width=1\textwidth]{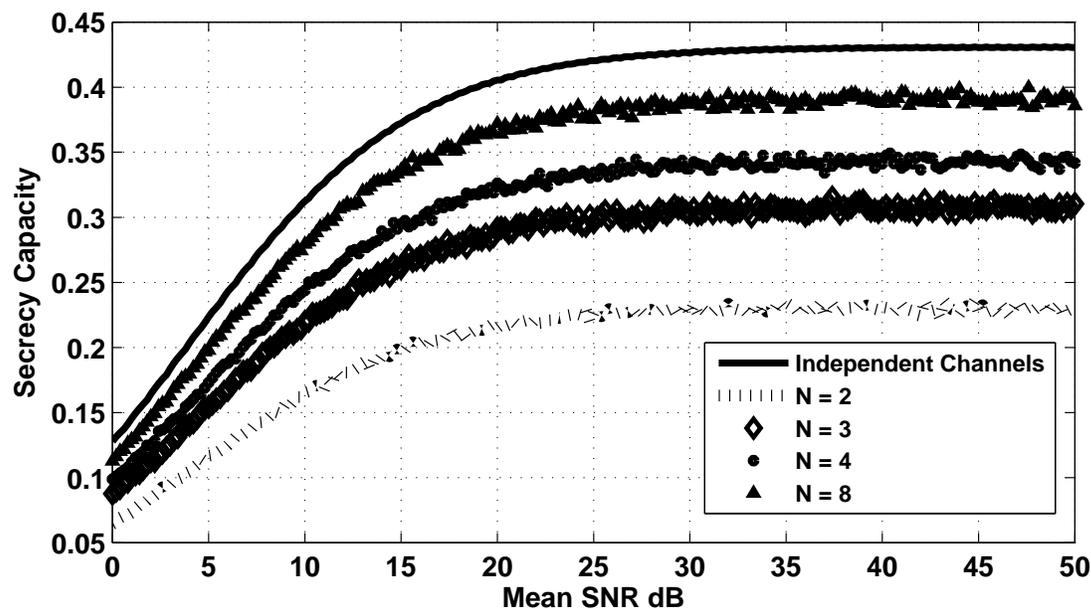}
\caption{The key rates using $N = 2,3,4,8$ dumb antennas, assuming fully correlated exponential channel gains. \label{fig5_dumb}}
\end{figure}

\begin{figure}
\centering
\includegraphics[width=1\textwidth]{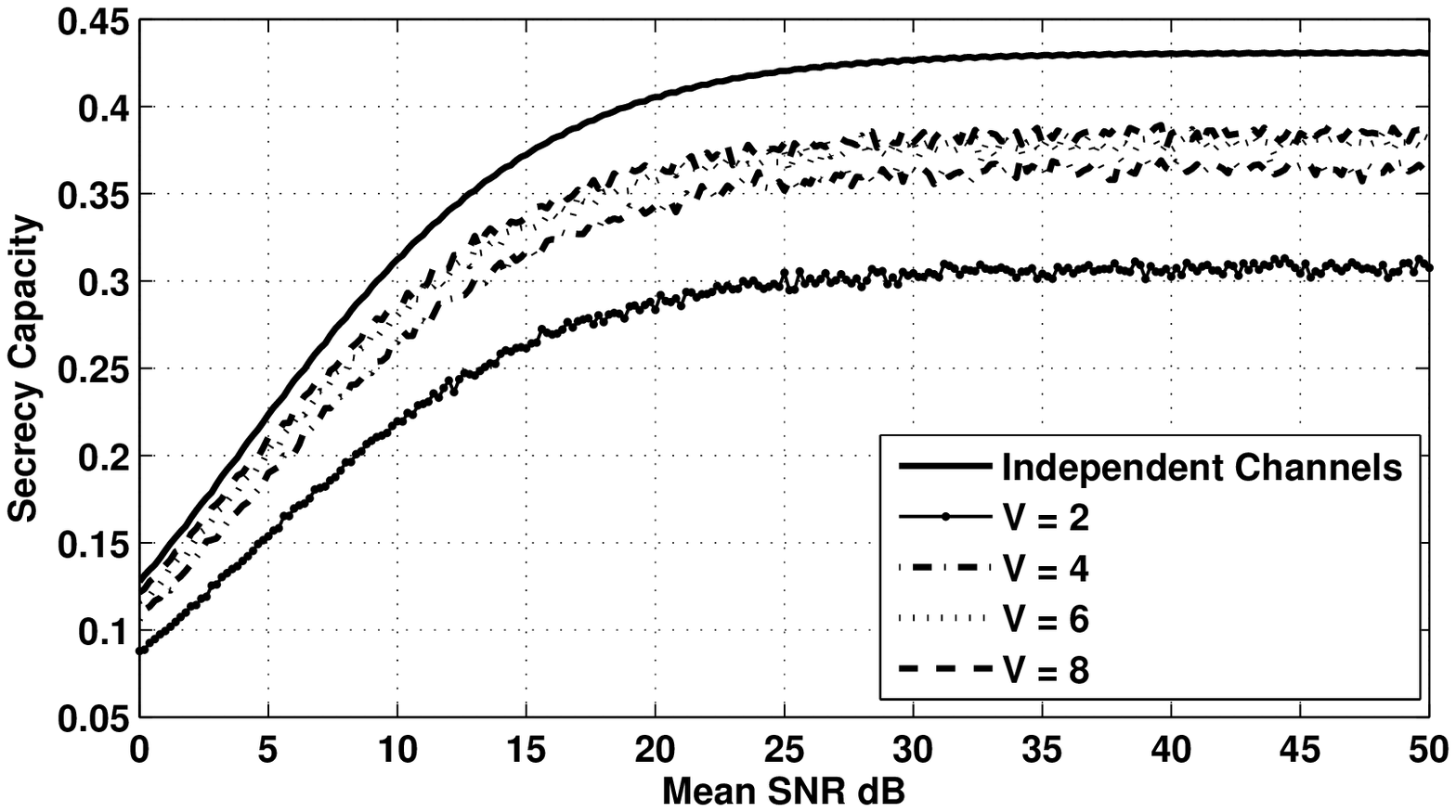}
\caption{The key rates using $N = 3$ dumb antennas, assuming fully correlated Chi-Square channel gains with different degrees of freedom $V = 2,4,6,8$. \label{fig6_dumb}}
\end{figure}

\begin{figure}
\centering
\includegraphics[width=1\textwidth]{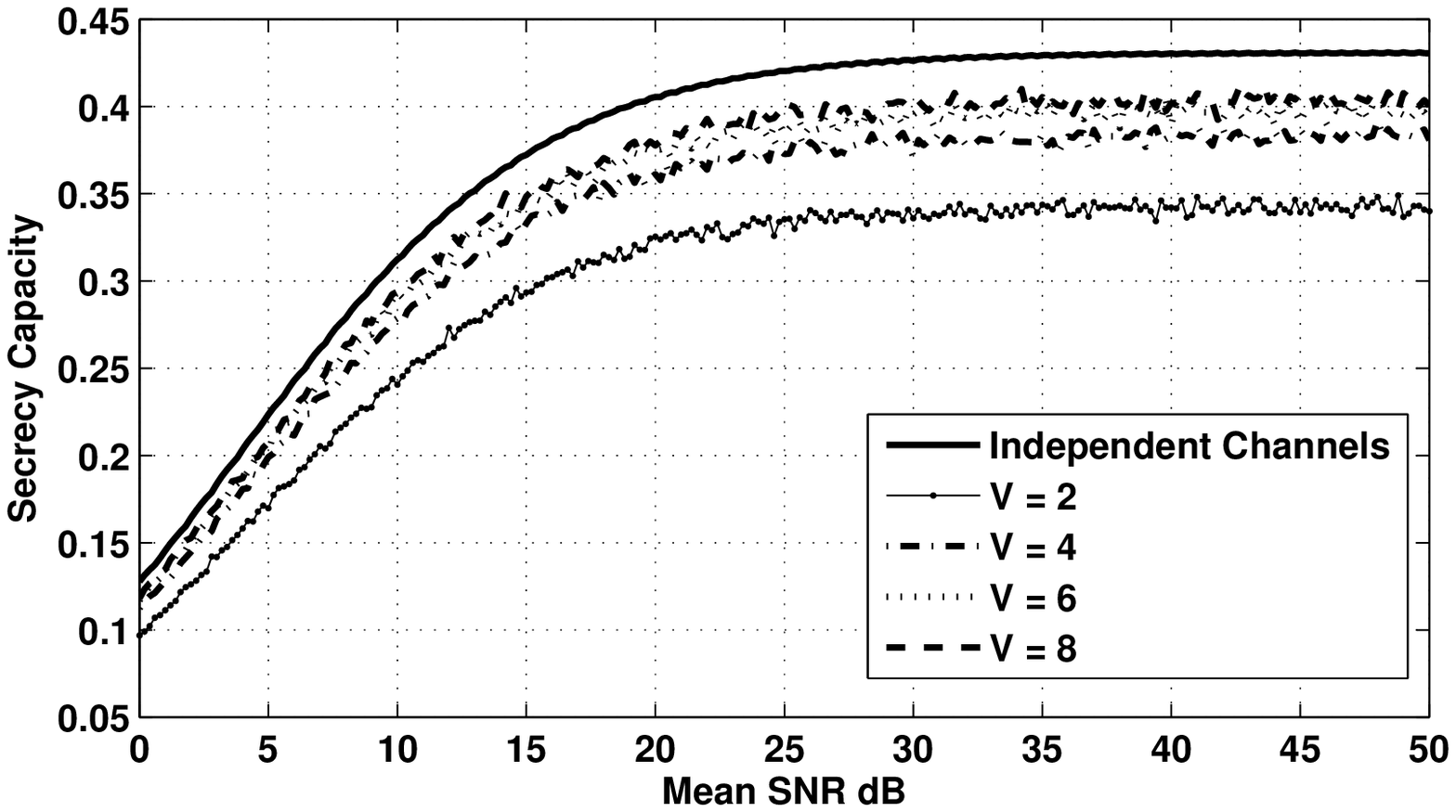}
\caption{The key rates using $N = 4$ dumb antennas, assuming fully correlated Chi-Square channel gains with different degrees of freedom $V = 2,4,6,8$. \label{fig7_dumb}}
\end{figure}

\begin{figure}
\centering
\includegraphics[width=1\textwidth]{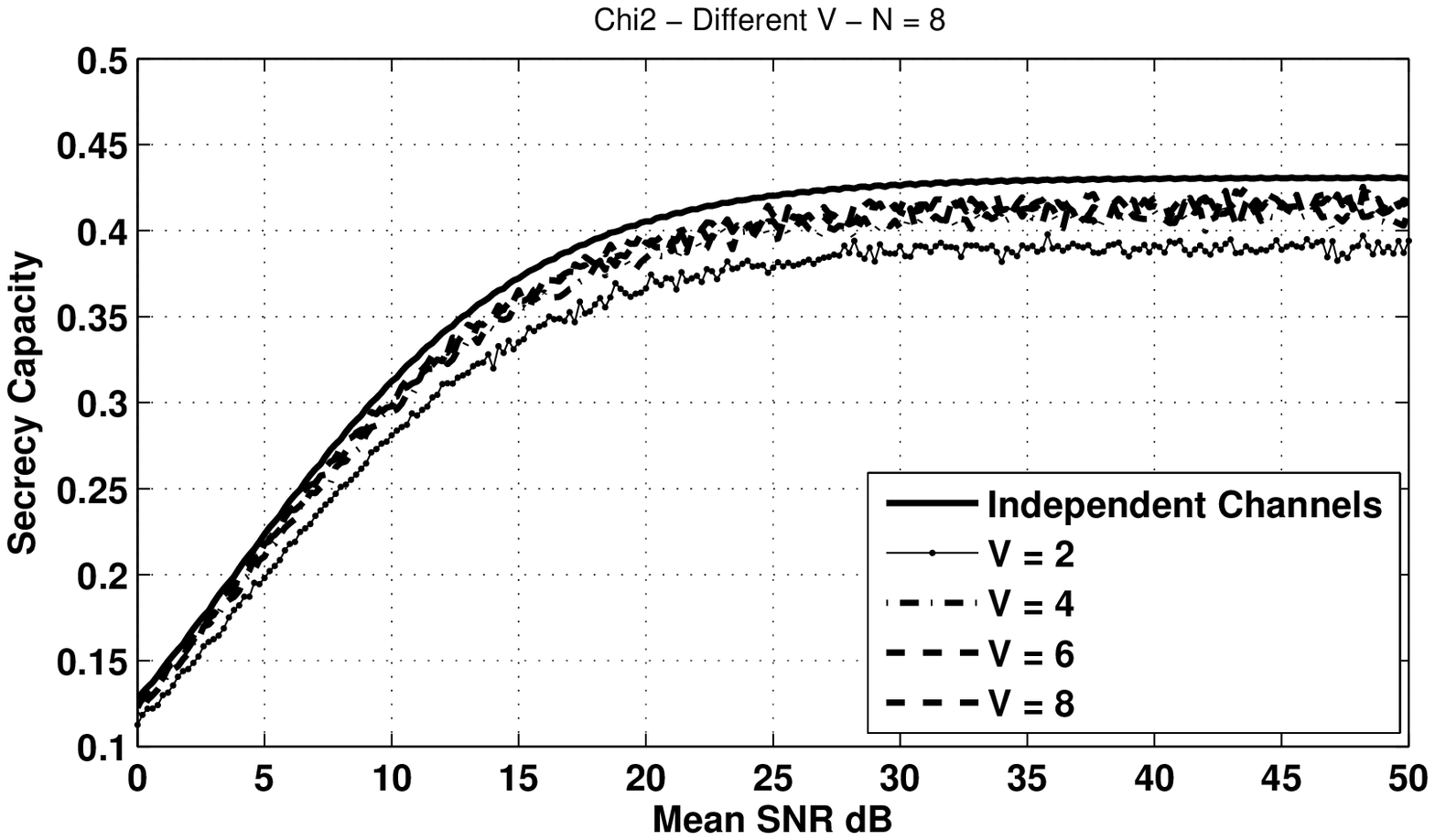}
\caption{The key rates using $N = 8$ dumb antennas, assuming fully correlated Chi-Square channel gains with different degrees of freedom $V = 2,4,6,8$. \label{fig8_dumb}}
\end{figure}

The role of dumb antennas in increasing the secrecy capacity of spatially correlated ARQ channels is investigated in the next set of figures. In our simulations, we assume that the channel gains are fully correlated, but the channel phases are independent. The independence assumption for the phases is justified as a small change in distance between Bob and Eve in the order of several electromagnetic wavelengths translates to a significant change in phase. Under these assumptions, it is easy to see that with one transmit antenna the secrecy capacity is zero. In Figure~\ref{fig5_dumb}, it is shown that as the number of antennas $N$ increases, the secret key rate approaches the upper bound given by~(\ref{sec_cap_special}) which assumes that the main and eavesdropper channels are independent. The same trend is observed in Figures~\ref{fig6_dumb},~\ref{fig7_dumb}, and~\ref{fig8_dumb} which generate the channel gains using chi-square distribution with different degrees of freedom. Overall, this set of results validates the theoretical claim of Appendix~\ref{proof_decorr}, indicating that dumb antennas can be used to de-correlate the main  and eavesdropper channels, even for a relatively small number of transmit antennas.

\begin{figure}
\centering
\includegraphics[width=1\textwidth]{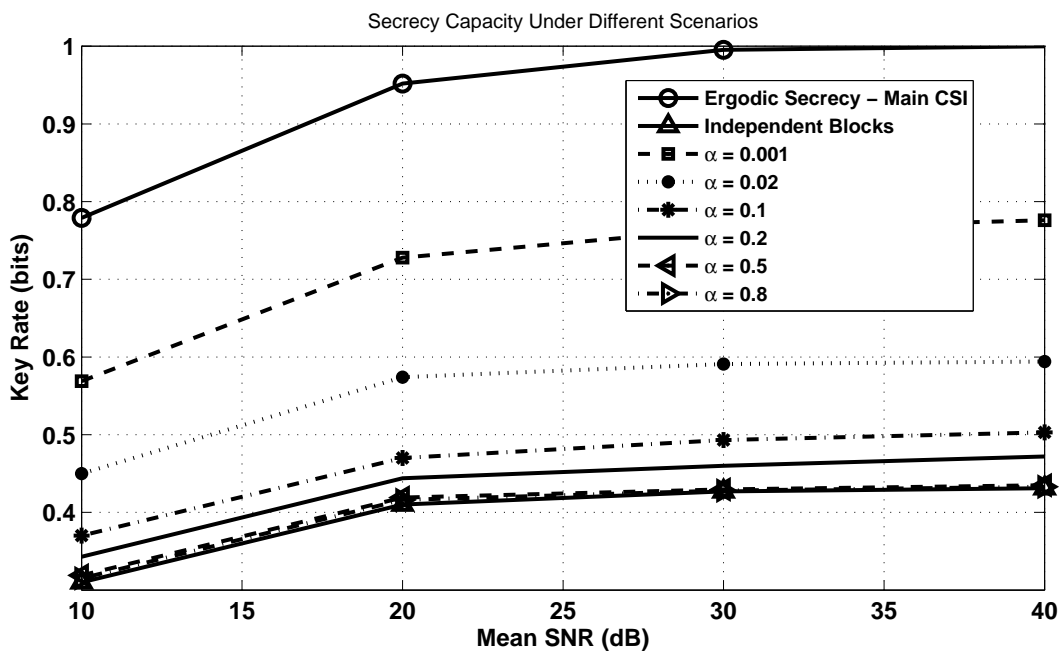}
\caption{The achievable key rates using the greedy scheme under different temporal correlation coefficient $\alpha$. \label{fig9_temp_corr}}
\end{figure}

Figure~\ref{fig9_temp_corr} reports the performance of the greedy rate adaptation algorithm for temporally correlated channels. The channel is assumed to follow the first order Markov model:
\begin{equation}
g(t) = (1 - \alpha) g(t-1) + \sqrt{2\alpha - \alpha^2} w(t)
\end{equation}
where $w(t)$ is the innovation process following $\mathcal{CN}(0,1)$ distribution. As expected, it is shown that as $\alpha$ decreases, the key rate increases. For the extreme points when $\alpha = 0$ or $\alpha = 1$, we get an \textbf{upper bound}, which is the ergodic secrecy under the main-channel transmit CSI assumption, and a \textbf{lower bound}, which is the ARQ secrecy capacity in case of independent block fading channel, respectively.

\section{Conclusions}\label{sec:conc}

This paper develops a novel {\bf overlay} approach for sharing secret keys using existing ARQ protocols. The underlying idea is to distribute the key bits over multiple ARQ frames and then use the authenticated ACK/NACK feedback to create an equivalent degraded channel at the eavesdropper. Our results establish the achievability of non-zero secrecy rates even when the eavesdropper is experiencing a higher average SNR than the legitimate receiver and shed light on the structure of optimal ARQ secrecy protocols. It is worth noting that our approach does not assume any prior knowledge about the instantaneous CSI; only prior knowledge of the average SNRs seen by the eavesdropper and the legitimate receiver are needed. Inspired by our information theoretic analysis, we have constructed low complexity secrecy coding schemes by transforming our channel to an erasure wiretap channel which lends itself to explicit coset coding approaches. Our secrecy capacity characterization reveals the negative impact of spatial correlation and the positive impact of temporal correlation on the achievable key rates. The former phenomenon is mitigated via a novel ``dumb antennas" technique, whereas the latter is exploited via a greedy rate adaptation policy. Finally, our theoretical claims have been validated via numerical examples that demonstrate the efficiency of the proposed
schemes.
The most interesting part of our work is, perhaps, the demonstration of the possibility of sharing secret keys
in wireless networks via rather simple modifications of the existing infrastructure which, in our case, corresponds to the ARQ mechanism. This observation motivated our follow-up work on developing secrecy protocols for Wi-Fi networks~\cite{arq-wep}.
\appendices

\section{Proof of Theorem~\ref{thm1}}\label{proof1}
In this appendix, we are going to prove both the achievability and converse of~(\ref{sec_cap_general}).

\subsection{Achievability Proof}
The proof is given for a fixed average power $P\leq
\bar{P}$ and transmission rate $R_0$. The key rate is then obtained
by the appropriate maximization. Let $R_s = C_s^{(g)} - \delta$ for some
small $\delta >0$ and $R = R_0 - \epsilon$. We first generate all binary sequences $\{
{\mathbf V} \}$ of length $m R$ and then independently assign each
of them randomly to one of $2^{n R_s}$ groups, according to a
uniform distribution. This ensures that any of the sequences are
equally likely to be within any of the groups. Each secret message
$w \in \{1, \cdots, 2^{n R_s} \}$ is then assigned a group ${\mathbf
V}(w)$. We then generate a Gaussian codebook consisting of $2^{n_1
\left(R_0 - \epsilon \right)}$ codewords, each of length $n_1$
symbols. The codebooks are then revealed to Alice, Bob, and Eve. To
transmit the codeword, Alice first selects a random group ${\mathbf
v}(i)$ of $n_1R$ bits, and then transmits the corresponding
codeword, drawn from the chosen Gaussian codebook. If Alice receives
an ACK bit from Bob, both are going to store this group of bits
and selects another group of bits to send in the next coherence
interval in the same manner. If a NACK was received, this group
of bits is discarded and another is generated in the same manner.
This process is repeated till both Alice and Bob have shared the
same key $w$ corresponding to $nR_s$ bits. We observe that the
channel coding theorem implies the existence of a Gaussian codebook
where the fraction of successfully decoded frames is given by
\begin{equation}
\frac{m}{n}=\textrm{Pr}(R_0 \leq \log_2(1+h_bP)),
\end{equation}
as $n_1\rightarrow\infty$. The equivocation rate at the eavesdropper can then be lower bounded as follows.
\begin{eqnarray}
n R_e &=& H(W|Z^n,K^b,G_b^b,G_e^b) \no\\
&\overset{(a)}{=}& H(W|Z^m,G_b^a,G_e^a) \no \\
&=& H(W,Z^m|G_b^a,G_e^a) - H(Z^m|G_b^a,G_e^a) \no\\
&=& H(W,Z^m,X^m|G_b^a,G_e^a) - H(Z^m|G_b^a,G_e^a) - H(X^m|W,Z^m,G_b^a,G_e^a) \no\\
&=& H(X^m|G_b^a,G_e^a) + H(W,Z^m| X^m,G_b^a,G_e^a) - H(Z^m|G_b^a,G_e^a) - H(X^m|W,Z^m,G_b^a,G_e^a) \no\\
&\ge& H(X^m|G_b^a,G_e^a) + H(Z^m| X^m,G_b^a,G_e^a) - H(Z^m|G_b^a,G_e^a) - H(X^m|W,Z^m,G_b^a,G_e^a) \no\\
&=& H(X^m|G_b^a,G_e^a) - I(Z^m ; X^m|G_b^a,G_e^a) - H(X^m|W,Z^m,G_b^a,G_e^a)\no
\end{eqnarray}
\begin{eqnarray}
&=& H(X^m | Z^m,G_b^a,G_e^a) - H(X^m|W,Z^m,G_b^a,G_e^a)\no\\
&\overset{(b)}{=}& \sum_{j=1}^a H(X(j)|Z(j),G_b(j),G_e(j)) - H(X^m|W,Z^m,G_b^a,G_e^a)\no \\
&\overset{(c)}{\ge}& \sum_{j \in {\mathcal N}_m} H(X(j)|Z(j),G_b(j),G_e(j)) - H(X^m|W,Z^m,G_b^a,G_e^a)\no \\
&=& \sum_{j \in {\mathcal N}_m} [ H(X(j)|G_b(j),G_e(j)) - I(X(j);Z(j)|G_b(j),G_e(j)) ] - H(X^m| W,Z^m,G_b^a,G_e^a) \no \\
&\ge & \sum_{j \in {\mathcal N}_m} n_1 \left[ R_0 - \log_2 \left( 1 + h_e(j) P \right) - \epsilon \right] - H(X^m|W,Z^m,G_b^a,G_e^a) \no \\
&\ge& \sum_{j=1}^{a} n_1 \left\{ \left[ R_0 - \log_2\left( 1 + h_e(j) P \right) \right]^{+} - \epsilon \right\} - H(X^m|W,Z^m,G_b^a,G_e^a) \no \\
&\overset{(d)}{=}& n C_s^{(g)} - H(X^m|W,Z^m,G_b^a,G_e^a) - m \epsilon.
\label{lb1}
\end{eqnarray}
In the above derivation, (a) results from the independent choice of
the codeword symbols transmitted in each ARQ frame which does not
allow Eve to benefit from the observations corresponding to the
NACKed frames, (b) follows from the memoryless property of the
channel and the independence of the $X(j)$'s, (c) is obtained by
removing all those terms which correspond to the coherence intervals
$j \notin {\mathcal N}_m$, where ${\mathcal N}_m = \left\{ j \in
\{1, \cdots,a\} : h_b(j) > h_e(j) | \psi = 1\right\}$, where $\psi$ is a binary random variable and $\psi = 1$ indicates that an ACK was received, and (d) follows from
the ergodicity of the channel as $n, m \rightarrow \infty$. Now we
show that the term $H(X^m|W,Z^m,G_b^a,G_e^a)$ vanishes as ${n_1} \to
\infty$ by using a list decoding argument. In this list decoding, at
coherence interval $j$, the wiretapper first constructs a list
${\mathcal L}_j$ such that ${\bf x}(j) \in {\mathcal L}_j$ if $({\bf
x}(i),{\bf z}(i))$ are jointly typical. Let ${\mathcal L}={\mathcal
L}_1 \times{\mathcal L}_2\times\cdots\times{\mathcal L}_a$. Given
$w$, the wiretapper declares that $\hat{{\bf x}}^m=({\bf x}^m)$ was
transmitted, if $\hat{x}^m$ is the only codeword such that
$\hat{{\bf x}}^m \in B(w)\cap {\mathcal L}$, where $B(w)$ is the set
of codewords corresponding to the message $w$. If the wiretapper
finds none or more than one such sequence, then it declares an
error. Hence, there are two types of error events: 1) ${\mathcal
E}_1$: the transmitted codeword ${\bf x}^m_t$ is not in ${\mathcal
L}$, 2) ${\mathcal E}_2$: $\exists {\bf x}^m \neq {\bf x}^m_t$ such
that ${\bf x}^m \in B(w)\cap{\mathcal L}$. Thus the error
probability $\Prob (\hat{{\bf x}}^m \neq {\bf x}^m_t )= \Prob (
{\mathcal E}_1\cup {\mathcal E}_2 ) \leq \Prob ( {\mathcal E}_1) +
\Prob ({\mathcal E}_2)$. Based on the Asymptotic Equipartition
Property (AEP)~\cite{Cover-IT}, we know that $\Prob ({\mathcal E}_1) \leq
\epsilon_1$. In order to bound $\Prob ({\mathcal E}_2)$, we first
bound the size of ${\mathcal L}_j$. We let
\begin{eqnarray}
\phi_j({\bf x}(j)|{\bf z}(j))=\left\{\begin{array}{ll}1,&
\textrm{$({\bf x}(j),{\bf z}(j))$ are jointly typical,} \\ 0,&
\textrm{otherwise.}\end{array} \right.
\end{eqnarray}
Now
\begin{eqnarray}
{\mathbb E}\{\|{\mathcal L}_j\|\}&=&{\mathbb E}\left\{\sum\limits_{{\bf x}(j)}\phi_j({\bf x}(j)|{\bf z}(j))\right\}\no\\
&\leq&{\mathbb E}\left\{1+\sum\limits_{{\bf x}(j) \neq {\bf x}_t(j)}
\phi_j({\bf x}(j)|{\bf z}(j))\right\}\no\\
&\leq&1+\sum\limits_{{\bf x}(j) \neq {\bf x}_t(j)}{\mathbb E}\left\{\phi_j({\bf x}(j)|
{\bf z}(j))\right\} \no\\
&\leq&1+2^{{n_1}\left[R_0 - \log_2(1+h_e(j)P)-\epsilon\right]}\no
\end{eqnarray}
\begin{equation}
{\mathbb E}\{\|{\mathcal L}_j\|\}\leq 2^{{n_1}\left(\left[R_0-\log_2(1+h_e(j)P) - \epsilon \right]^+
+\frac{1}{n_1} \right)}
\end{equation}
Hence
\begin{eqnarray}
{\mathbb E}\{\|{\mathcal L}\|\}&=&\prod\limits_{j=1}^{a} {\mathbb
e}\{\|{\mathcal L}_j\|\}=2^{\sum\limits_{j=1}^a n_1\left(\left[R_0-
\log_2(1+h_E(j)P) - \epsilon \right]^+ + \frac{1}{n_1}\right) }\\
\Prob ({\mathcal E}_2 ) &\leq& {\mathbb E}\left\{\sum\limits_{x^m
\in{\mathcal L}, {\bf x}^m \neq {\bf x}^m_t}
\Prob ({\bf x}^m \in B(w)) \right\}\no\\
&\overset{(a)}\leq& {\mathbb E}\left\{\|{\mathcal
L}\|2^{-n R_s}\right\}\no \\
&\leq& 2^{-n R_s}2^{\sum\limits_{j=1}^a
n_1\left(\left[R_0-\log_2(1+h_e(j)P) - \epsilon \right]^+ +
\frac{1}{n_1} \right)
}\no\\
&\leq& 2^{-n \left(R_s -\frac{1}{c}\sum\limits_{j=1}^a
\left(\left[R_0
-\log_2(1+h_e(j)P) - \epsilon \right]^+ + \frac{1}{n_1} \right) \right)}, \no \\
&=& 2^{-n \left(R_s -\frac{1}{c}\sum\limits_{j=1}^a \left(\left[R_0
-\log_2(1+h_e(j)P) \right]^+ + \frac{1}{n_1} \right) +
\frac{|{\mathcal N}_m| \epsilon}{c} \right)},
\end{eqnarray}
where (a) follows from the uniform distribution of the codewords in
$B(w)$. Now as $n_1 \to \infty$ and $a \to \infty$, we get \[ \Prob
({\mathcal E}_2 ) ~\le ~ 2^{-n \left(C_s^{(g)} - \delta - C_s^{(g)} + a \epsilon
\right)} ~=~ 2^{-n(c \epsilon -\delta)}, \] where $c = \Prob ( h_b >
h_e)$. Thus, by choosing $\epsilon > (\delta / c)$, the error
probability $\Prob ({\mathcal E}_2 ) \to 0$ as $n \to \infty$. Now
using Fano's inequality, we get \[ H(X^m|W,Z^m,G_b^a,G_e^a) ~\leq~ n
\delta_{n} \qquad \mbox{$\to 0$} \qquad \mbox{as  $m,n \to \infty$}.
\] Combining this with (\ref{lb1}), we get the desired result.\\

\subsection{Converse Proof}
We now prove the converse part by showing that for any perfect secrecy rate $R_s$ with equivocation rate $R_e > R_s - \epsilon$ as $n,m \rightarrow \infty$, there exists a transmission rate $R_0$, such that $$R_s ~\leq~ \mathbb{E}\left\{\left[R_0 - \log_2\left(1+h_eP\right)\right]^{+}\mathbb{I}\left(R_0 \leq \log_2\left(1+h_bP\right) \right)\right\} $$ Consider any sequence of $(2^{nR_s},m)$ codes with perfect secrecy rate $R_s$ and equivocation rate $R_e$, such that $R_e > R_s - \epsilon$ as $n \rightarrow \infty$.
We note that the equivocation $H(W|Z^n,K^n,G_b^b,G_e^b)$ only depends on the marginal distribution of $Z^n$, and thus does not depend on whether $Z(i)$ is a physically or stochastically degraded version of $Y(i)$ or vice versa. Hence we assume in the following derivation that for any fading state, either $Z(i)$ is a physically degraded version of $Y(i)$ or vice versa (since the noise processes are Gaussian). Thus we have
\begin{eqnarray}
nR_e &=& H(W|Z^b,K^n,G_b^b,G_e^b) \no \\
&\overset{(a)}=& H(W|Z^m,G_b^a,G_e^a) \no\\
&\overset{(b)}{\le}& H(W|Z^m,G_b^a,G_e^a) - H(W|Z^m,Y^m,G_b^a,G_e^a) + m\delta_m \no\\
&=& I(W;Y^m|Z^m,G_b^a,G_e^a) + m\delta_n \no
\end{eqnarray}
\begin{eqnarray}
&\overset{(c)}{\le}&  I(X^m;Y^m|Z^m,G_b^a,G_e^a) + m\delta_m \no \\
&=& H(Y^m|Z^m,G_b^a,G_e^a) - H(Y^m|X^m,Z^m,G_b^a,G_e^a) + m\delta_m \no \\
&=& \sum_{i=1}^a [ H(Y(i)|Y^{i-1},Z^m,G_b^a,G_e^a) - H(Y(i)|Y^{i-1},X^m,Z^m,G_b^a,G_e^a) ] + m\delta_m \no \\
&\overset{(d)}{\le}& \sum_{i=1}^a [ H(Y(i)|Z(i),G_b(i),G_e(i) - H(Y(i)|X(i),Z(i),G_b(i),G_e(i)) ] + m\delta_m \no \\
&=& \sum_{i=1}^a I(X(i);Y(i)|Z(i),G_b(i),G_e(i)) + m\delta_m \no \\
&\overset{(e)}{=}& \sum_{i=1}^a I(X(i);Y(i)|G_b(i),G_e(i)) - I(X(i);Z(i)|G_b(i),G_e(i)) + m\delta_m \no \\
&{\le}&  \sum_{i=1}^a R_0 - \log_2(1+h_e(i)P) + m\delta_m \no \\
&{\le}&  \sum_{i=1}^a [ R_0 - \log_2(1+h_e(i)P) ]^+ + m\delta_m \no \\
R_e &\overset{(f)}{\le}& \mathbb{E}\left\{\left[R_0 - \log_2\left(1+h_eP\right)\right]^{+}\mathbb{I}\left(R_0 \leq \log_2\left(1+h_bP\right) \right)\right\} + \beta\delta_m
\end{eqnarray}where $\beta = \textrm{Pr}(R_0 \leq \log_2(1+h_bP))$\\
In the above derivation, (a) results from the independent choice of the codeword symbols transmitted in each ARQ frame which does not allow Eve to benefit from the observations corresponding to the NACKed frames, (b) follows from Fano's inequality, (c) follows from the data processing inequality since $W \to X^m \to (Y^m,Z^m)$ forms a Markov chain, (d) follows from the fact that conditioning reduces entropy and from the memoryless property of the channel, (e) follows from the fact that $I(X;Y|Z) = I(X;Y) - I(X;Z)$ as shown in~\cite{Wyner1}, (f) follows from ergodicity of the channel as $m,n \rightarrow \infty$. The claim is thus proved.

\section{Proof of Decorrelation}\label{proof_decorr}
In this appendix, we show that employing multiple transmit antennas makes the correlation between Eve's and Bob's channel power gains converge to zero, in a mean-square sense, as the number of antennas $N$ goes to $\infty$. Let
\begin{equation}
l_1 = |g_b^{eq}|^2 \hspace{0.1in}\textrm{and}\hspace{0.1in} l_2 = |g_e^{eq}|^2\no
\end{equation}
Assuming all $\theta$'s to be uniformly distributed in the interval $[-\pi,\pi]$, we get,
\begin{eqnarray}
l_1 &=& \frac{1}{N}\left[\left\vert\sum\limits_{i=1}^{N}\cos\left(\theta_{iR}+\theta_{iB}\right)\right\vert^2 + \left\vert\sum\limits_{i=1}^{N}\sin\left(\theta_{iR}+\theta_{iB}\right)\right\vert^2\right]\no\\
&=& \frac{1}{N}\left[N + 2\sum\limits_{i=1}^{N-1}\sum\limits_{j=i+1}^{N}\left\{\cos\left(\theta_{iR}+\theta_{iB}\right)\cos\left(\theta_{jR}+\theta_{jB}\right) + \sin\left(\theta_{iR}+\theta_{iB}\right)\sin\left(\theta_{jR}+\theta_{jB}\right)\right\}\right]\no\\
&=& 1 + \frac{2}{N} \sum\limits_{i=1}^{N-1}\sum\limits_{j=i+1}^{N}\cos\left(\theta_{iR}+\theta_{iB}-\theta_{jR}-\theta_{jB}\right)\label{l1eq}
\end{eqnarray}

\noindent Similarly for $l_2$,
\begin{equation}
l_2 = 1 + \frac{2}{N} \sum\limits_{i=1}^{N-1}\sum\limits_{j=i+1}^{N}\cos\left(\theta_{iR}+\theta_{iE}-\theta_{jR}-\theta_{jE}\right)
\label{l2eq}
\end{equation}
Now, taking the expectation of~(\ref{l1eq}) and~(\ref{l2eq}) with respect to the random phases applied on the transmit antenna array $\theta_{iR}$ for given values of $\theta_{iE}$'s and $\theta_{iB}$'s, we get,
\begin{eqnarray}
\mathbb{E}\left(l_1\right) &=& \mathbb{E}\left(l_2\right) = 1\\
\mathbb{E}\left(l_1l_2\right) &=& 1 + \frac{2}{N^2} \sum\limits_{i=1}^{N-1}\sum\limits_{j=i+1}^{N}\cos\left[\left(\theta_{iB}-\theta_{iE}\right) - \left(\theta_{jB}-\theta_{jE}\right)\right]\\
\mathbb{E}\left(l_1^2\right) &=& \mathbb{E}\left(l_2^2\right) = 1 + \frac{2}{N^2}\frac{N(N-1)}{2} = 1 + \frac{N-1}{N}
\end{eqnarray}
So, the variance of $l_1$ and $l_2$ is given by,
\begin{equation}
\textrm{var}\left(l_1\right) = \textrm{var}\left(l_2\right) = \sigma^2_{l_1} = \sigma^2_{l_2} = \frac{N-1}{N}
\end{equation}
Therefore, the correlation coefficient $\rho$ between the channels' power gains is given by
\begin{eqnarray}
\rho &=& \frac{\mathbb{E}\left(l_1l_2\right) - \mathbb{E}\left(l_1\right)\mathbb{E}\left(l_2\right)}{\sqrt{var\left(l_1\right)}\sqrt{Var\left(l_2\right)}}\no\\
&=& \frac{2}{N(N-1)} \sum\limits_{i=1}^{N-1}\sum\limits_{j=i+1}^{N}\cos\left[\left(\theta_{iB}-\theta_{iE}\right) - \left(\theta_{jB}-\theta_{jE}\right)\right]\no\\
&=& \frac{2}{N(N-1)} \sum\limits_{i=1}^{N-1}\sum\limits_{j=i+1}^{N}\cos\left[\Delta_i - \Delta_j\right]
\end{eqnarray}
where
\begin{equation}
\Delta_i = \theta_{iB}-\theta_{iE}\hspace{0.1in} \textrm{and} \hspace{0.1in}\Delta_j = \theta_{jB}-\theta_{jE}
\end{equation}
Assuming $\theta_{iB}, \theta_{iE}, \theta_{jB}, \theta_{jE}$ are all independent, and uniformly distributed in the interval $[-\pi,\pi]$, and taking the expectation of $\rho$ over them, we get,
\begin{equation}
\mathbb{E}\left(\rho\right) = 0
\end{equation}
The divergence of $\rho$ around its mean is given by,
\begin{eqnarray}
\textrm{var}(\rho) &=& \sigma^2 = \frac{4}{N^2(N-1)^2}\sum\limits_{i=1}^{N-1}\sum\limits_{j=i+1}^{N}\textrm{var}\left(\cos\left(\Delta_i - \Delta_j\right)\right)\no\\
&=& \frac{4}{N^2(N-1)^2}.\frac{N(N-1)}{2}.\frac{1}{2}\no\\
&=& \frac{1}{N(N-1)}\label{corr_conv}
\end{eqnarray}

\noindent Thus, the standard deviation of $\rho$ is given by:
\begin{equation}
\sigma = \frac{1}{\sqrt{N(N-1)}} \simeq \frac{1}{N}
\end{equation}

\noindent It is evident from~(\ref{corr_conv}) that $\textrm{var}(\rho)$ goes to zero as $N \rightarrow \infty$. That is, the correlation coefficient $\rho$ converges, in a mean-square sense, to zero.


\ifCLASSOPTIONcaptionsoff
  \newpage
\fi



%

\newpage

\end{document}